\documentclass[aps,pra,twocolumn,showpacs,showkeys]{revtex4}
\usepackage{epsfig}
\usepackage{hyperref}

\begin{document}

\def\il{I_{low}} 
\def\iu{I_{up}} 
\def\eeq{\end{equation}}
\def\ie{i.e.}  
\def\etal{{\it et al. }}
\def\pr{Phys. Rev. }  
\def\prb{Phys. Rev. {\bf B}}
\def\pra{Phys. Rev. {\bf A}} 
\def\prl{Phys. Rev. Lett. }
\def\pla{Phys. Lett. A } 
\def\pb{Physica B}
\def\ajp{Am. J. Phys. } 
\def\jpc{J. Phys. C } 
\def\rmp{Rev. of Mod. Phys. } 
\def\jap{J. Appl. Phys. } 
\def\mpl{Mod. Phys. Lett. {\bf B}} 
\def\ijmp{Int. J. Mod. Phys. {\bf B}} 
\def\ijp{Ind. J. Phys. }
\def\ijpap{Ind. J. Pure Appl. Phys. }
\def\ibmjrd{IBM J. Res. Dev. }
\def\pjp{Pramana J. Phys.}

\title{Wave attenuation to clock sojourn times.}
\author{Colin Benjamin}
\email{colin@iopb.res.in}
\author{A. M. Jayannavar}
\email{jayan@iopb.res.in}
\affiliation{Institute of Physics, Sachivalaya Marg, Bhubaneswar 751 005,
  Orissa, India}
\date{\today}

\begin{abstract}
  The subject of time in quantum mechanics is of perennial interest
  especially because there is no observable for the time taken by a
  particle to transmit (or reflect) from a particular region.  Several
  methods have been proposed based on scattering phase shifts and
  using different quantum clocks, where the time taken is clocked by
  some external input or indirectly from the phase of the scattering
  amplitudes. In this work we give a general method for calculating
  conditional sojourn times based on wave attenuation. In this
  approach clock mechanism does not couple to the Hamiltonian of the
  system. For simplicity, specific case of a delta dimer is considered
  in detail. Our analysis re-affirms recent results based on
  correcting quantum clocks using optical potential methods, albeit in
  a much simpler way.
   
\end{abstract}

\maketitle

There has been considerable interest on the question of time spent by
a particle (interaction time) in a scattering region or in a given
region of space\cite{landauer,hauge,gasparian}.  This problem has been
approached from many different points of view, but there is no clear
consensus about a simple expression for this time as there is no
hermitian operator associated with it (although experimentalists have
claimed to measure it\cite{landauer}). The prospect of nanoscale
electronic devices has in recent years brought new urgency to this
problem as this is directly related to the maximum attainable speed of
such devices.  When it comes to quantum phenomena of tunneling, the
time taken by a particle to traverse the barrier is a subject of
controversy till now\cite{landauer,hauge}. In some formulations this
time leads to a real quantity and in others to a complex
quantity\cite{baskin}. In certain cases tunneling time is considered
to be ill-defined or quantum mechanics does not allow us to discuss
this time\cite{baskin,yamada,steinberg}. Furthermore sometimes it is
maintained that tunneling through a barrier takes zero
time\cite{reflandauernature}. Recently, Anantha Ramakrishna and Kumar
(AK)\cite{anantha,ananthathesis} have proposed the non unitary Optical
potential as a clock to calculate the sojourn times without the clock
affecting it. In this paper we examine another non-unitary clock,
i.e.,wave attenuation (or, stochastic absorption) to calculate the
sojourn lengths, i.e., the total effective distance travelled by a
particle in the region of interest. This sojourn length on appropriate
division by the speed of the particle in the region of interest will
give us the sojourn time.

Before, we go to the main body of our paper it would be appropriate to
describe the relative merits and demerits of the proposals up-till
now.  The first proposal was by Wigner\cite{wigner} and the so called
Wigner delay time defined separately for reflection or transmission.
These are asymptotic times and include self interference delays as
well as the time spent in the barrier. This method is based on
following the peak of the wave packet, and it loses it's significance
under strong distortion of the wave packet\cite{landauer}. Moreover,
there is no causal relationship between the peak of the transmitted
packet and the peak of the incident packet. This is due to the fact
that peak of the transmitted packet can leave the scattering region
long before the peak of the incident packet has arrived. In this
treatment one cannot address the time spent by a particle in a local
region of scattering.  However, it should be emphasized that this time
is of physical importance as it is intimately connected to the dynamic
admittance and other physical properties of
microstructures\cite{3joshithesis}. The next proposal of dwell
time\cite{smith,but} is an exact statement of the time spent in a
region of space averaged over all incoming particles. The problem with
this is that it does not distinguish between reflection and
transmission and consequently it cannot answer the question ``How much
time a transmitted (alternatively, Reflected) particle spent in the
scattering region?''. Some other proposals \cite{lanbut,but} include
the oscillating barrier which considers only a particular limit, i.e.,
the opaque barrier limit, and others invoke a physical
clock\cite{but,butchamber} possesing extra degrees of freedom that
co-evolves with the sojourning particle. In some of these treatments
the sum of the times spent in two-non overlapping intervals is
non-additive and also the very clock mechanism affects the sojourn
time to be clocked finitely even as the perturbing clock potential is
infinitesimally small\cite{anantha,ananthathesis}. This raises the
important question namely ``Can quantum mechanical sojourn time be
clocked without the clock affecting it?''. To this end AK have
proposed a non-unitary clock wherein the absorption caused by an
infinitesimal optical potential formally introduced over the locality
of interest acts as a physical clock to count the time of sojourn in
it. The problem for this non-unitary clock is that the optical
potential itself introduces spurious scattering's and this affects the
time to be clocked. In a novel manner AK propose a formal procedure by
which these spurious scattering's are eliminated and the sojourn times
clocked by this optical potential counter satisfies all the necessary
conditions especially it is real, additive and distinguishes between
reflection and transmission.
 
In this paper we introduce the wave attenuation (or stochastic
absorption) method to calculate the sojourn lengths and times. In this
method\cite{datta,joshi} we damp the wave function by adding an
exponential factor ($e^{-\alpha l}$) every time we traverse the length
of interest, here $2\alpha$ represents the attenuation per unit
length.  This method is better than the optical potential model as it
does not suffer from spurious scattering's\cite{CB,jayan,rubio}. The
corrections introduced in case of Optical potential model to take care
of spurious scattering's will become manifestly difficult when we
calculate the sojourn times for a superlattice involving numerous
scatterer's. Thus our method of wave attenuation scores over the
optical potential model. Moreover, this method will be helpful in
calculating delay times in presence of incoherence which have been
done earlier using imaginary potentials.

In the presence of wave attenuation a wave attenuates exponentially
and thus the transmission (or reflection) coefficient becomes
exponential with the length endured in presence of the attenuator and
this acts as a natural counter for the sojourn length.  Following the
procedure of AK we calculate the sojourn lengths and times in case of
propagation as outlined in (Fig.~1).
\begin{figure}[h]
\protect\centerline{\epsfxsize=3.5in \epsfbox{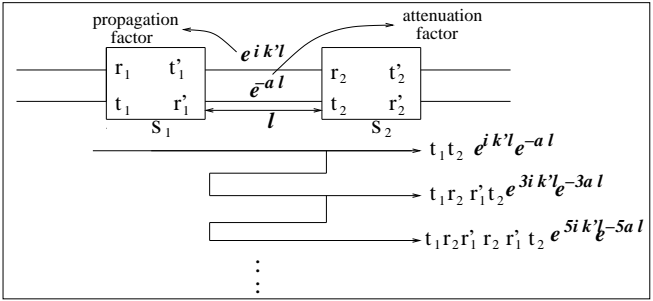}}
\caption{Summing the different paths, $S_1$ and $S_2$ denote the two
  scatterer's. $l$ is the distance between them. $e^{i
    k^{\prime}l}$ and $e^{-\alpha l}$ denote the propagation and attenuation
    factors in the locality of interest.}
\end{figure}

We calculate the closed form formulas for reflection and transmission
coefficients as also the sojourn lengths and times in case of
propagation. The amplitude for transmission and reflection can be
calculated by summing\cite{datta} the different paths as in Fig.~1.
The scatterer $S_{1}$ in Fig.~1 has as its elements
$r_{1}^{},r_{1}^{\prime},t_1^{}$ and $t_1^{\prime}$. $r_{1}^{}$ is the
reflection amplitude when a particle is reflected from the left side
of the barrier while $r_{1}^{\prime}$ is the reflection amplitude when
a particle is reflected from the right side of the barrier. $t_1^{}$
and $t_1^{\prime}$ are the amplitudes for transmission when a particle
is transmitted from left to right of the barrier and vice-versa.
Similar assignments are done for the scatterer $S_2^{}$.
\begin{figure}[h]
\protect\centerline{\epsfxsize=3.5in \epsfbox{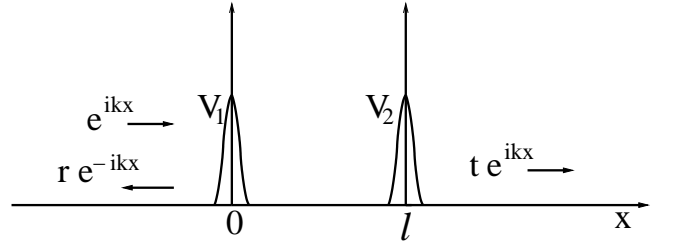}}
\caption{The delta dimer}
\end{figure}

Thus for the amplitude of transmission we have-
$t=t_{1}^{}t_{2}^{}e^{i k^{\prime}l}e^{-\alpha l}
+t_{1}^{}r_{2}^{}r^{\prime}_{1}t_{2}^{}e^{3 i k^{\prime}l}e^{-3 \alpha
  l} +...$ which can be summed as
\begin{eqnarray}
t=\frac{t_{1}^{}t_{2^{}}e^{i k^{\prime}l}e^{-\alpha
           l}}{1-r_{2}^{}r^{\prime}_{1}e^{2 i k^{\prime}l}e^{-2 \alpha l} }
\end{eqnarray}
and this is the transmission amplitude in presence of wave
attenuation.  Again for the case of reflection amplitude we have
$r=r_{1}^{}+t_{1}^{}r_{2}^{}t^{\prime}_{1}e^{2 i k^{\prime}l}e^{-2 \alpha
  l}+t_{1}^{}r_{2}^{}r^{\prime}_{1}r_{2}^{}t^{\prime}_{1}e^{4 i
  k^{\prime}l}e^{-4 \alpha l}+
t_{1}^{}r_{2}^{}r^{\prime}_{1}r_{2}^{}r^{\prime}_{1}r_{2}^{}t^{\prime}_{1}e^{6
  i k^{\prime}l}e^{-6 \alpha l}+.. $ , which leads to -
\begin{eqnarray}
r=r_{1}^{}+ \frac{t_{1}^{}r_{2}^{}t^{\prime}_{1}e^{2 i
    k^{\prime}l}e^{-2 \alpha
           l}}{1-r^{\prime}_{1}r_{2}^{}e^{2 i k^{\prime}l}e^{-2 \alpha l} }
\end{eqnarray}
or 
\begin{eqnarray}
r=\frac{r_{1}^{}-a r_{2}^{}e^{2i k^{\prime}l}e^{-2\alpha
           l}}{1-r^{\prime}_{1}r_{2}^{}e^{2 i k^{\prime}l}e^{-2 \alpha l} }
\end{eqnarray}

In Eq.~3, $a=r_{1}^{}r^{\prime}_{1}-t_{1}^{}t^{\prime}_{1}$ is the
determinant of the S-Matrix of the first barrier and as we are only
dealing with unitary S-Matrices therefore the determinant is of
unit modulus for all barriers.  In these expressions
$k^{\prime}$ is the wave vector in the region of interest.  The
transmission and reflection coefficients can be calculated by taking
the square of the modulus of the expressions in Eq's.~(1) and (3).

The sojourn lengths for transmission and reflection are calculated as
below\cite{anantha}-
The sojourn length for transmission is given as 
\begin{eqnarray}
l^{T}=\lim_{2\alpha \rightarrow 0} -\frac{\partial (\ln
  |t|^2)}{\partial (2 \alpha)} 
\end{eqnarray}
and for reflection is defined as- 
\begin{eqnarray}
l^{R}=\lim_{2\alpha \rightarrow 0} - \frac{\partial (\ln
  |r|^2)}{\partial (2 \alpha)} 
\end{eqnarray}
The sojourn times for reflection or transmission can be calculated
from the formula- $\tau^{R/T}=\frac{l^{R/T}}{|\frac{\hbar
    k^{\prime}}{m}|}$ wherein as before $|\frac{\hbar k^{\prime}}{m}|$
is the speed of propagation in the region of interest.  From Eq's.~(4)
and (5) we can calculate the sojourn lengths for transmission-
\begin{eqnarray}
\frac{l^{T}}{l}=\frac{1-|r^{\prime}_{1}|^{2}|r_{2}^{}|^{2}}{1-2
  \Re(r^{\prime}_{1}r_{2}^{} e^{2 i k^{\prime} l}) +|r^{\prime}_{1}|^{2}|r_{2}^{}|^{2}}
\end{eqnarray}
and for reflection-
\begin{eqnarray}
\frac{l^{R}}{l}=\frac{l^{T}}{l}+\frac{|r_{2}^{}|^{2}-|r_{1}^{}|^{2}}{|r_{1}^{}|^{2}-2
  \Re(r^{*}_{1}r_{2}^{}a  e^{2 i k^{\prime} l}) +|r_{2}^{}|^{2}}
\end{eqnarray}
Here $\Re$ represents real part of the quantity in brackets.
In the above two equations the sojourn lengths have been normalized
with respect to the length $l$ of the locality of interest. Throughout 
the discussion the quantities are expressed in their dimensionless
form.
\begin{figure}[h]
\protect\centerline{\epsfxsize=3.5in \epsfbox{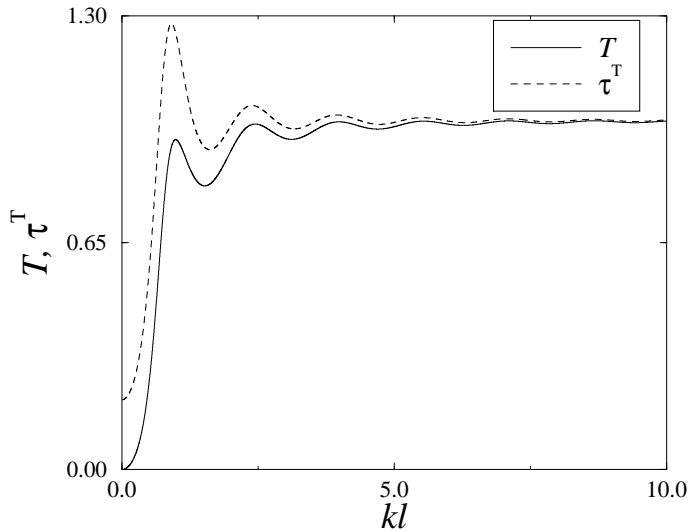}}
\caption{T and $\tau^{T}$ for a non symmetric delta
  dimer. $l=2.0$, $V_{1}$=1.0 and  $V_{2}=0.5.$}
\end{figure}

\begin{figure}[h]
\protect\centerline{\epsfxsize=3.5in \epsfbox{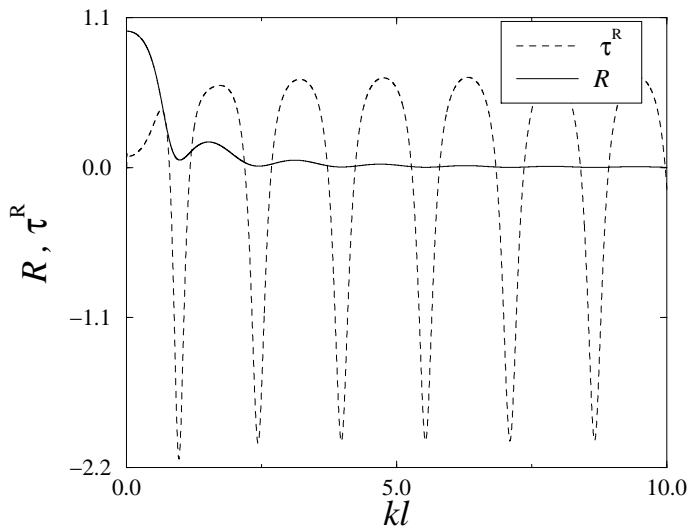}}
\caption{R and $\tau^{R}$ for a non symmetric delta
  dimer. $l=2.0$, $V_{1}^{}$=1.0 and  $V_{2}^{}=0.5.$ }
\end{figure}
 
We consider the case of a delta dimer with strength $V_1$ and $V_2$ as
shown in Fig.~2 separated by a distance $l$.  The potentials
$V_j^{}$'s in their dimensionless form are expressed as $V_j
\equiv\frac{V_j l}{k^{\prime}\hbar^{2}/2m}, j=1,2$.

The S-Matrices for the delta dimer are given as-
\[S_j=\left(\begin{array}{cc}
    r_j^{} & t_j^{\prime}\\
    t_j^{} & r_j^{\prime}\\
\end{array} \right) \ =\left(\begin{array}{cc}
\frac{V_j}{2ik-V_j}     & \frac{2ik}{2ik-V_j} \\
\frac{2ik}{2ik-V_j}     & \frac{V_j}{2ik-V_j} \\
\end{array} \right)\]  
In Figures~(3) and (4) we plot the Transmission and Reflection
coefficients and the normalized sojourn times $\tau^{T}$ and
$\tau^{R}$ for a non symmetrical delta dimer. We have normalized these
sojourn times by the time taken by a particle to traverse a distance
$l$ in absence of barrier, i.e., $\frac{m l}{|\hbar k^{\prime}|}$.
From Eq.~6 it is clear that transmission time is always positive
definite.  Moreover we can readily show from our treatment that local
times spent by the particle in non-overlapping regions are additive.
In Fig.~3 we have plotted the transmission and the normalized sojourn
time $\tau^{T}$ for a non symmetrical delta dimer. As expected the
sojourn times are larger near the resonances.  When transmission
approaches unity in the large $kl$ limit, the dimensionless
transmission sojourn time approaches unity. In Fig.~4 we have plotted
the reflection coefficient and the normalized sojourn time $\tau^{R}$.
We observe that for an unsymmetrical delta-dimer the sojourn time in
case of reflection goes negative for certain values of the potentials.
Hence we get negative sojourn times in certain regions for the case of
reflection. Even though additivity of local sojourn times in two
non-overlapping regions holds here too.  It has been argued by AK that
this is because in case of reflection there is a partial wave
corresponding to prompt reflection $r_{1}$ that never samples the
region of interest, and also this prompt part leads to self
interference delays which cause the sojourn time $\tau^{R}$ to become
negative for some values of the parameters. If one removes this prompt
part as suggested by AK, i.e., $r_{np}=r-r_{1}^{}$, and we calculate
the sojourn time $\tau^{R_{np}}$ with this prompt part removed we find
it to be positive definite and given by $\tau^{R_{np}}=\tau^{T}+1$, as
$\tau^{T}$ is positive definite. Removing the prompt part of course
cures the problem of negative sojourn times but it causes another
problem we find when $V_{2}$ is zero, $\tau_{R}=0$, as expected
because particle is not reflected after entering the region as there
is no barrier to the right of the locality of interest, but
$\tau^{R_{np}}=2$. Even in absence of barriers when there is no
reflection still we get $\tau^{R_{np}}=2$, which follows trivially
from the formal final expression $\tau^{R_{np}}=\tau^{T}+1$, as
$\tau^T=1$ irrespective of the case whether $V_2^{}=0$ or both
$V_1^{}$ and $V_2^{}$ are zero. This is unexpected looking at the
expression for $r$ in Eq.~3. This leads us to the conclusion that to
get a positive definite answer for the reflection time by removing
prompt part requires careful thinking. In case of a symmetric barrier
as is obvious from Eq's.~6 and 7 $\tau^T=\tau^R$. Again in case of a
non-symmetric delta dimer $\tau^T$ is independent of the fact whether
particle is incident from the right or left, but $\tau^R$ depends
non-trivially on the direction of incidence of the particle.
  
A remarkable assertion found in the literature\cite{hauge} concerning the
measurement of the time of transmission or reflection is $\tau^{D}=T
\tau^{T}+R \tau^{R}$. Herein $\tau^{T}$ and $\tau^{R}$ are as given
above while the dwell time $\tau^{D}=\frac{1}{v}\int_{0}^{l} |\psi|^2
dx$. $\psi$ is the wavefunction in the locality of interest and $v$ is
the speed of the particle in the region of interest.  In Fig.~ 5 we
plot $\tau^{D}$ and $ T \tau^{T}+R \tau^{R}$, for a symmetrical delta
dimer. We find that they are inequivalent in the small $kl$ regime
\cite{landauer}.

\begin{figure}[h]
\protect\centerline{\epsfxsize=3.5in \epsfbox{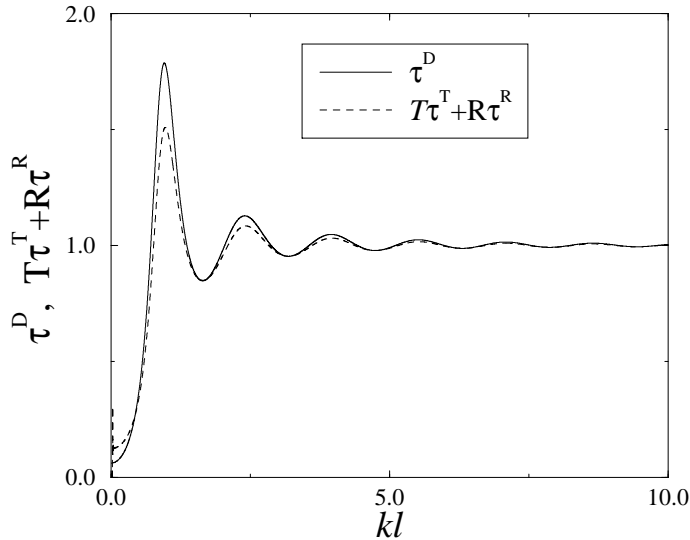}}
\caption{$\tau^{D}$ and $ T \tau^{T}+R \tau^{R}$,
  are inequivalent. The parameters are $l=2.0$, $V_{1}=1.0$ and $V_{2}=1.0$.}
\end{figure}

For the case of rectangular barriers of height $V$, $r$ and $t$'s are
the reflection and transmission amplitudes at the interfaces and
$k^{\prime}=\sqrt\frac{2 m (E-V)}{\hbar^2}$ in the propagation regime and
the speed is $|\frac{\hbar k^{\prime}}{m}|$. For this case we get the
same result as obtained in the Ref.\cite{anantha}.

In conclusion, we have given a simple method of calculating sojourn
times using wave attenuation method. In this method, wave attenuation
(or, stochastic absorption) as we have treated cannot be incorporated
in a Hamiltonian. Thus it inherently takes care of spurious
scattering's, which arise when Hamiltonians with an optical potential
are used (or for every clock where perturbation due to clock mechanism
couples to the Hamiltonian\cite{ananthathesis}).  Our results reaffirm
those obtained by AK after taking care of the spurious contributions
from the optical potential. The transmission sojourn times are always
positive definite and are additive as mentioned above. Reflection
sojourn times can become negative. By removal of the prompt part of
reflection $r_{1}$ one gets positive reflection sojourn time but it
further leads to another problem. Thus, if we insist on reflection
sojourn time to be positive then it requires further careful analysis,
or, reflection time itself needs closer inspection.  Finally when we
compare the dwell time with the conditional sojourn times weighted
over appropriate reflection and transmission coefficients, we find
them not to be equivalent.

\begin{acknowledgments}
  The authors would like to thank Dr. S. Anantha Ramakrishna for useful 
  discussions on this problem and one of us (AMJ) thanks
  Professor N. Kumar for his continual interest and discussions on
  this problem over the years. 
\end{acknowledgments}

\end{document}